\documentclass[twocolumn,english,prl]{revtex4-1}
\usepackage[T1]{fontenc}
\usepackage[latin9]{inputenc}
\usepackage{amsmath}
\usepackage{amssymb}
\usepackage{graphicx}

\makeatletter
\@ifundefined{date}{}{\date{}}
\usepackage{braket}

\usepackage{babel}

\makeatother

\usepackage{babel}
\begin{document}
\title{Nonlocal pair correlations in a higher-order Bose gas soliton}
\author{King Lun Ng and Bogdan Opanchuk }
\affiliation{Centre for Quantum and Optical Science, Swinburne University of Technology,
Melbourne 3122, Australia}
\author{Margaret D. Reid and Peter D. Drummond}
\affiliation{Centre for Quantum and Optical Science, Swinburne University of Technology,
Melbourne 3122, Australia}
\affiliation{Institute of Theoretical Atomic, Molecular and Optical Physics (ITAMP),
Harvard University, Cambridge, Massachusetts, USA. }
\begin{abstract}
The truncated Wigner and positive-P phase-space representations are
used to study the dynamics of a one-dimensional Bose gas. This allows
calculations of the breathing quantum dynamics of higher-order solitons
with $10^{3}-10^{5}$ particles, as in realistic Bose-Einstein condensation
(BEC) experiments. Although classically stable, these decay quantum
mechanically. Our calculations show that there are large nonlocal
correlations. These also violate the Cauchy-Schwarz inequality, showing
the presence of nonclassical quantum entanglement. 
\end{abstract}
\maketitle
A classical soliton \cite{Russell1844,Korteweg_deVries1895} is a
non-dispersive pulse caused by the balance of dispersion and nonlinearity
in a nonlinear wave. Their initial shape can be maintained during
propagation. Higher-order classical solitons have additional spatiotemporal
oscillations. Time invariant quantum solitons can exist \cite{mcguire1964study,thacker1981exact},
but are completely delocalized in phase and in space. When a coherent
soliton is prepared that is classically invariant, quantum effects
change the soliton shape. These have been theoretically predicted
\cite{Carter1987,Drummond1993,haus1990quantum} and experimentally
verified \cite{rosenbluh1991squeezed,Drummond1993-solitons,spalter1998observation,Corney_2008}.

Higher-order solitons have attracted much recent interest, since their
quantum fluctuations can become macroscopically large \cite{WernerPhysRevA.54.R2567,Streltsov2008,cosme2016center,yurovsky2017dissociation,Opanchuk2017_1Dsoliton}.
This leads to a macroscopic quantum initiated decay with fragmentation
into multiple condensates, reminiscent of the decay of a false vacuum
in scalar quantum field theory \cite{coleman1977fate}. In this Letter,
we show that these quantum effects are accompanied by nonlocal dynamical
correlations, which occur even before the soliton decays. These fluctuations
are largest for a Bose-Einstein condensate soliton formed at mesoscopic
particle number. This may be testable in proposed experiments \cite{yurovsky2017dissociation}
in bosonic $^{7}Li$, with $10^{3}-10^{4}$ Bose condensed atoms.
These correlations survive to very large particle number, and we obtain
measurable predictions even for small density changes.

One-dimensional attractive Bose gases form a bright soliton in photonic
\cite{mollenauer1980experimental} and Bose-Einstein condensate environments
\cite{Khaykovich2002,Strecker2002,nguyen2017formation}. In treatments
of these Bose gases, the classical description is known in optics
as the nonlinear Schrödinger equation, and in atomic physics as the
Gross-Pitaevskii equation \cite{Gross1963,Gross1961,Pitaevskii1961}.
This equation uses a mean-field approximation such that operator products
are assumed to factorize. In order to include the full quantum properties,
beyond mean-field methods are required that include quantum correlations,
allowing predictions of rich quantum features. These provide tests
of many-body quantum dynamics in a highly controlled, experimentally
accessible environment.

An early prediction in photonic systems was the generation of quantum
squeezing and entanglement \cite{Carter1987,Drummond1993} in one-dimensional
bright solitons, verified experimentally in photonic experiments \cite{Drummond1993-solitons,Corney_2008,rosenbluh1991squeezed}.
More recently, there has been interest in the quantum dynamical evolution
of higher-order solitons, which oscillate periodically in space at
the mean-field level. They can be generated from a fundamental soliton
with a \emph{sech} envelope by a rapid increase in the coupling constant.
In atomic gases, this is obtainable through a Feshbach resonance.
Ultra-cold atomic physics allows for a strong coupling regime, with
fewer particles than in photonics.

Here we use quantum phase space methods to analyze this quench experiment,
in which the full many-body quantum state is sampled probabilistically,
allowing a calculation of the dynamical evolution of nonlocal correlation
functions. These are known to be good indicators of entanglement and
possible Bell violations in BEC systems \cite{kheruntsyan2012violation,bonneau2018characterizing,wasak2018bell,reid1986violations}.
The main technique used is a truncated Wigner (tW) method \cite{Wigner1932}
that employs a $1/N$ expansion for $N$ particles, with $N=10^{3}-10^{5}$,
as in currently proposed experiments. The general approach has been
verified through accurate predictions of quantum squeezing in optical
fibre solitons \cite{Carter1987,Drummond1993,Corney:2006_ManyBodyQD}.
All the local conservation laws of the bright BEC soliton system are
preserved \cite{Drummond2017_TWD}. We also confirm these results
using the exact positive-P phase-space representation \cite{Drummond1980}
up to the first oscillation peak.

The density dynamics, but not correlations, have been calculated previously.
An approximate variational prediction \cite{Streltsov2008} using
the many-body multi-configurational time-dependent Hartree for bosons
method (MCTDHB), predicted a sudden fragmentation into two equal fragments.
Later work \cite{cosme2016center}, pointed out that this MCTDHB approximation
failed to predict the known center-of-mass variance growth. This is
because the calculation used only two modes, while there are seven
or more condensate modes present \cite{Opanchuk2017_1Dsoliton}. Other
methods using exact eigenstates \cite{yurovsky2017dissociation} or
the DMRG approximation \cite{weiss2016higher}, have several orders
of magnitude fewer particles.

For Bose gases strongly confined in a one-dimensional waveguide along
the $r$ direction with a transverse trapping frequency $\omega_{\perp}$,
the Hamiltonian in the occupation number representation is given by
\begin{eqnarray}
\hat{H}_{1D} & = & \int dr\left[\frac{-\hbar^{2}}{2m}\hat{\Psi}^{\dagger}\frac{\partial^{2}}{\partial r^{2}}\hat{\Psi}+\frac{g}{2}\Psi^{\dagger2}\hat{\Psi}^{2}\right],\label{eq:H1d}
\end{eqnarray}
where $\hat{\Psi}\left(r\right)$ is a one-dimensional quantum field
operator.

The total many-body Hamiltonian includes two-body s-wave collisions
where $g=2\hbar a\omega_{\perp}$ is the interaction strength. This
is tunable, since the $s$-wave scattering length $a$ is a function
of the external magnetic field via a Feshbach resonance \cite{PollackHuletPhysRevA.81.053627}.
Taking a characteristic length scale $r_{0}$ and time scale $t_{0}$
where $r_{0}^{2}=\hbar t_{0}/2m$, the length and time are transformed
into dimensionless form{{} $z=r/r_{0}$ }and $\tau=t/t_{0}$. The
interaction strength $g$ is also transformed into a scaled quantity
$C=mgr_{0}/\hbar^{2}$. The corresponding Hamiltonian \cite{lieb1963exact,mcguire1964study}
for a system of dimensionless length $L$ is 
\begin{eqnarray}
\hat{h} & = & \int_{0}^{L}dz\left[-\hat{\psi}^{\dagger}(z)\nabla_{z}^{2}\hat{\psi}(z)+C\hat{\psi}^{\dagger2}(z)\hat{\psi}^{2}(z)\right].\label{cold-collision-1}
\end{eqnarray}

We assume an initial Poissonian distribution of particle numbers,
which is a good approximation to the lowest observed experimental
BEC number fluctuations in small condensates of $10^{3}$ particles
\cite{chuu2005direct}, and corresponds to a coherent state. In the
Wigner representation, the field operator $\hat{\psi}(z)$ is replaced
by a stochastic field $\psi$ \cite{Drummond1993,Steel1998,opanchuk2013functional},
which in a symmetrically ordered mapping is initially 
\begin{eqnarray}
\psi(z) & = & \sqrt{n_{0}(z)}+\frac{1}{\sqrt{2}}\underset{k}{\sum}\frac{1}{\sqrt{L}}\eta_{k}e^{ikz}.\label{eq:Wigner}
\end{eqnarray}
Here $\eta_{k}$ is a complex number with correlations $\langle\eta_{k}\eta_{k'}\rangle=0$
and $\langle\eta_{k}\eta_{k'}^{*}\rangle=\delta_{kk'}$, while $n_{0}(z)=\text{\ensuremath{\langle\hat{\psi}^{\dagger}}(z)\ensuremath{\hat{\psi}}(z)}\rangle$.
An alternative approach is to use the positive-P representation \cite{Drummond1980,Carter1987},
which is exact, equivalent to normal ordering, and has two stochastic
fields $\psi,\psi^{+}$ with initial values $\psi(z)=\psi^{+}(z)=\sqrt{n_{0}(z)}$.

The Bose gas is assumed to be initially trapped with a weakly attractive
interaction $C=-2/N$, with $n_{0}(z)=Nsech^{2}(z)/2$. At time $\tau=0$,
a rapid change of interaction strength is activated by turning on
a negative interaction $C=-8/N$. These parameters are chosen to be
the same as that of earlier studies \cite{Streltsov2008,cosme2016center,Opanchuk2017_1Dsoliton}.
This is equivalent to an experimental system of photons or atoms with
an interaction quench which increases the interaction strength by
a factor of $4$.

The resulting quantum dynamical equation of motion in the truncated
Wigner representation is 
\begin{eqnarray}
\frac{d\psi}{d\tau} & = & i\nabla^{2}\psi-2iC\psi\left(|\psi|^{2}-2\epsilon\right)+O(1/N),\label{eq:equation of motion}
\end{eqnarray}
where $\epsilon=1/2\Delta z$ is an ordering correction for a computational
lattice spacing of $\Delta z$. The $O(1/N)$ term represents higher-order
differential operators in the phase-space evolution equations, which
are neglected here.

The quantum time-evolution dynamical equations in the positive-P case
are \cite{Carter1987}: 
\begin{eqnarray}
{\frac{d\psi}{d\tau}} & {=} & {i\nabla_{z}^{2}\psi-2iC\psi^{+}\psi{}^{2}-i\sqrt{2iC}\psi\eta\left(\tau,z\right)}\label{eq:equation of motion-1}\\
{\frac{d\psi^{+}}{d\tau}} & {=} & {-i\nabla_{z}^{2}\psi+2iC\psi^{+2}\psi-\sqrt{2iC}\psi^{+}\eta^{+}\left(\tau,z\right),}\nonumber 
\end{eqnarray}
with independent complex Gaussian stochastic noises $\eta,\eta^{+}$,
having non-vanishing correlations: 
\begin{align}
\langle\eta\left(\tau,z\right)\eta\left(\tau',z'\right)\rangle & =\langle\eta^{+}\left(\tau,z\right)\eta^{+}\left(\tau',z'\right)\rangle\nonumber \\
 & =\delta\left(\tau-\tau'\right)\delta\left(z-z'\right).
\end{align}

The partial differential equations were solved using an interaction
picture fourth-order Runge-Kutta (RK4) method, using two different
public-domain software packages \cite{Kiesewetter2016xspde,Opanchuk2014-reikna},
with identical results in both cases. The results given here use the
truncated Wigner method, as it has much lower sampling error for long
times in this system. These were replicated up to the first oscillation
peak with the positive-P equations, as a check on these results.

\begin{figure}[h]
\includegraphics[width=0.9\columnwidth]{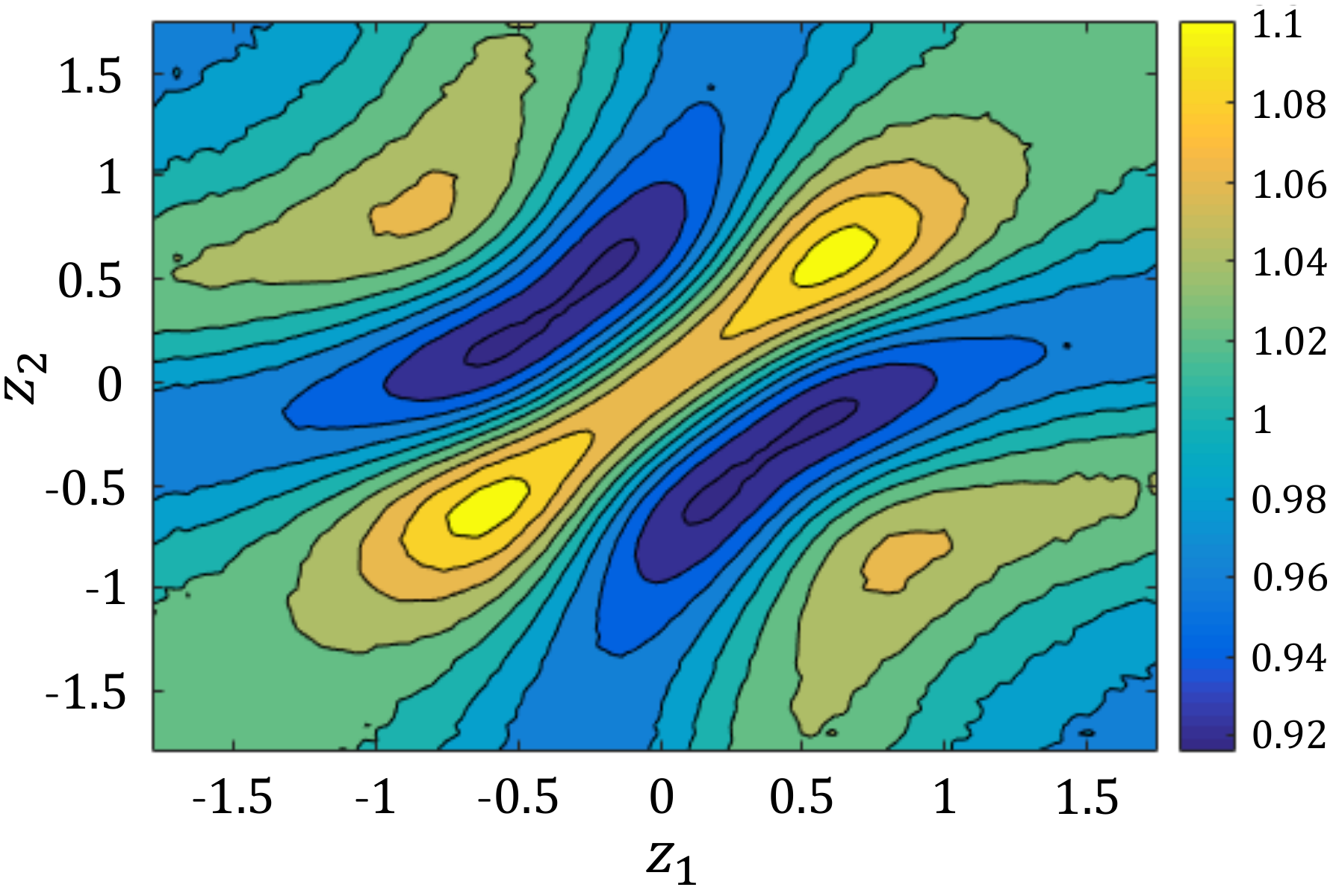}

\includegraphics[width=0.9\columnwidth]{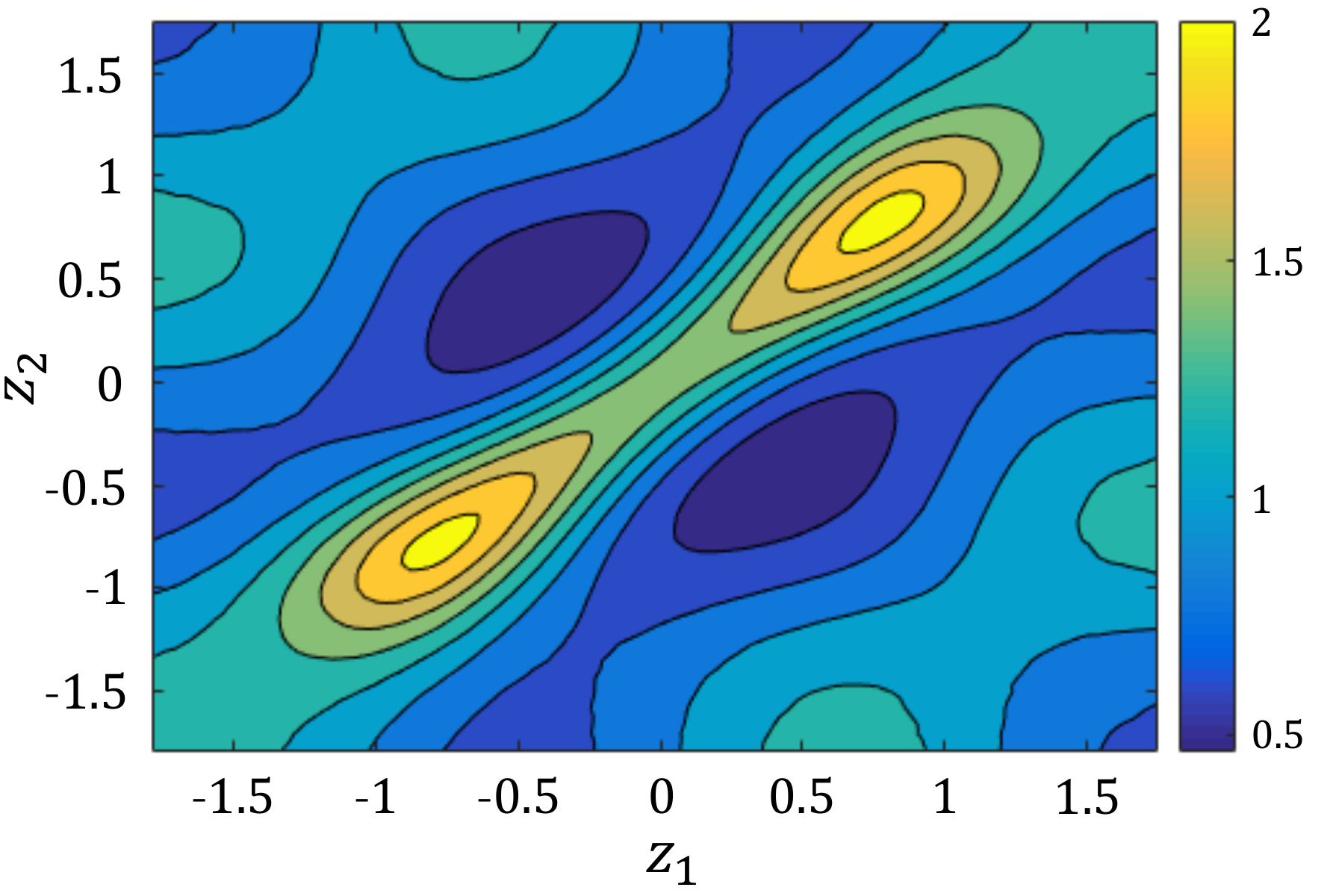}

\caption{\label{fig:g2 contour N1e3}Normalized second order correlation $g^{(2)}(z_{1},z_{2})$
for $N=10^{3}$ at $\tau=1.0$ (top) and $\tau=5.0$ (bottom). Here
$CN=-8$ and $L=20$. Simulations have $M=512$ modes, $9\times10^{4}$
trajectories and $9\times10^{4}$ time steps. The contour plots show
the region within $z=\pm1.78$. The diagonal axis shows the local
correlation $g^{(2)}(\bar{z},\bar{z})$, with $\bar{z}=\left(z_{1}+z_{2}\right)/2$.
The anti-diagonal axis shows the nonlocal correlation $g^{(2)}(z_{1},z_{2})$
where $z_{1}=-z_{2}$. The maximum sampling error is $10^{-2}$ and
the maximum time-step error is $10^{-6}$. }
\end{figure}
A similar calculation has been performed in \cite{Drummond2017_TWD,Opanchuk2017_1Dsoliton}.
This demonstrated that all four local conservation laws are satisfied
in the simulations. The time-evolution of the density of the classical
soliton system near the center ($z=0$) oscillates with constant period.
However, the true quantum condensate fragments into multiple smaller
Bose condensates. Thus, the soliton gradually breaks up due to quantum
effects.

Here we investigate the quantum correlations caused by this instability.
To do this we compare soliton experiments with different number of
particles $N$ while keeping $CN$ constant, so the classical results
are the same up to a scaling factor. Defining $n_{i}=n\left(z_{i}\right)\equiv|\psi(z_{i})|^{2}$,
the measurable quantum correlations are given by the second order
intensity correlation $G^{(2)}(z_{1},z_{2})=\langle\hat{\psi}^{\dagger}(z_{1})\hat{\psi}^{\dagger}(z_{2})\hat{\psi}(z_{2})\hat{\psi}(z_{1})\rangle$
\cite{glauber1963quantum} . In terms of the Wigner ensemble averages,
this is:{ 
\begin{equation}
G^{(2)}(z_{1},z_{2})=\text{\ensuremath{\langle}}n_{1}n_{2}-\epsilon\left(1+\delta_{z_{1}z_{2}}\right)\left[n_{1}+n_{2}-\epsilon\right]\rangle_{W}.
\end{equation}
}

The normalized correlation function is given by 
\begin{eqnarray}
g^{(2)}(z_{1},z_{2}) & = & \frac{\langle\hat{\psi}^{\dagger}(z_{1})\hat{\psi}^{\dagger}(z_{2})\hat{\psi}(z_{2})\hat{\psi}(z_{1})\rangle}{\langle\hat{n}(z_{1})\rangle\langle\hat{n}(z_{2})\rangle},\label{eq:g2 correlation}
\end{eqnarray}
where we note that the product of annihilation and creation operators
is expressed in terms of the Wigner representation, so that the expected
number density is: 
\begin{eqnarray}
\langle\hat{n}(z)\rangle=\langle\hat{\psi}^{\dagger}(z)\hat{\psi}(z)\rangle & = & \text{\ensuremath{\langle}}n(z)\rangle_{W}-\epsilon.\label{eq:wigner density}
\end{eqnarray}

The normalized correlation function is used to observe the bunching
$\left(g^{(2)}(z,z)>1\right)$ and anti-bunching $\left(g^{(2)}(z,z)<1\right)$
amplitude of the soliton. According to the contour plot displayed
in Figure \ref{fig:g2 contour N1e3}, for the $N=10^{3}$ system,
the 1D BEC soliton develops a strong bunching region with peaks of
$g^{(2)}$ increasing from $\sim1.1$ at $\tau=1.0$ (Figure \ref{fig:g2 contour N1e3}
top) to $\sim2$ at $\tau=5.0$ (Figure \ref{fig:g2 contour N1e3}).

\begin{figure}[h]
\includegraphics[width=0.9\columnwidth]{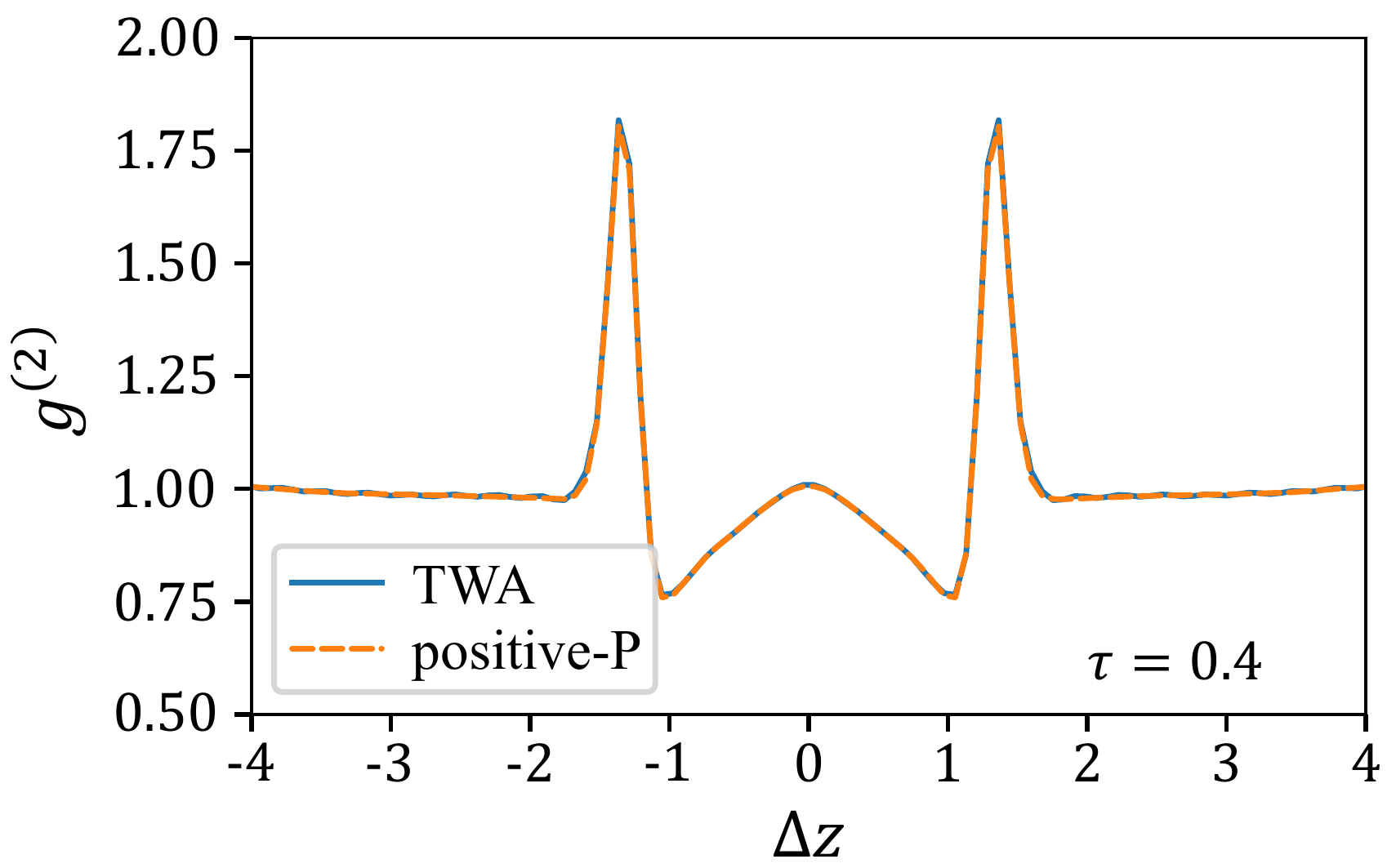}

\caption{\label{fig:g2 comparison-slice N1e3-1}Normalized second order correlation
$g^{(2)}(\Delta z)\equiv g^{(2)}(\Delta z/2,-\Delta z/2)$ for $N=10^{3}$,
$\tau=\pi/8$. Simulations have $M=512$ modes, $9\times10^{4}$ trajectories
and $9\times10^{4}$ time steps. Graphs compare calculations with
the truncated Wigner approximation and exact positive-P representations,
showing complete agreement within the width of the graphed lines.}
\end{figure}

When using the normally-ordered positive-P representation, the normally
ordered averages require no ordering corrections. In this case we
define $n_{i}\equiv\psi^{+}(z_{i})\psi(z_{i}),$ and one finds that
$G^{(2)}(z_{1},z_{2})=\text{\ensuremath{\langle}}n_{1}n_{2}\rangle_{P}$,
and $\langle\hat{n}(z)\rangle=\text{\ensuremath{\langle}}n(z)\rangle_{P}.$
This method has no $N$-dependent truncation, which allows us to confirm
that truncation errors are negligible in the Wigner predictions. We
find no difference in the results, as expected, given that $N\ge1000$
for these calculations. Figure \ref{fig:g2 comparison-slice N1e3-1}
shows complete agreement of the two simulations for $g^{(2)}(\Delta z)\equiv g^{(2)}(\Delta z/2,-\Delta z/2)$.
This gives nonlocal anti-correlations and correlations at the first
peak, occurring at $\tau=\pi/8$.

\begin{figure}[h]
\includegraphics[width=1\columnwidth]{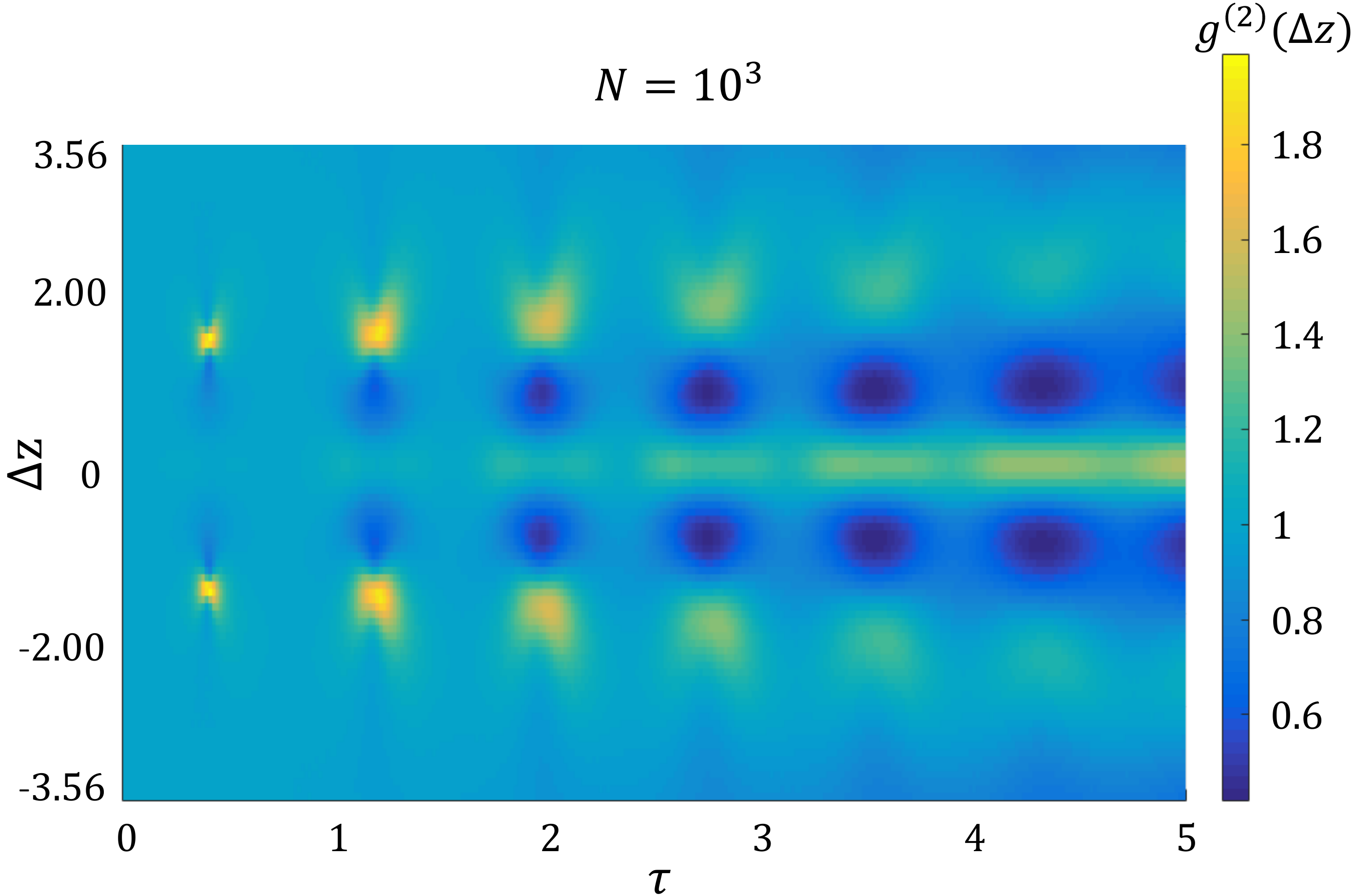}

\caption{\label{fig:g2 time evolution-10^3}Time-evolution of the normalized
second order correlation $g^{(2)}(\Delta z)$, where $N=10^{3}$,
$CN=-8$ and $L=20$. Simulations with $M=512$ modes, $9.6\times10^{4}$
trajectories and $9\times10^{4}$ time steps. These contours correspond
to the correlation along the anti-diagonal axis in Figure \ref{fig:g2 contour N1e3}
which represent the nonlocal correlation. The maximum sampling error
is around $0.2\%$ of $g^{(2)}(\Delta z)$. }
\end{figure}

\begin{figure}[h]
\includegraphics[width=1\columnwidth]{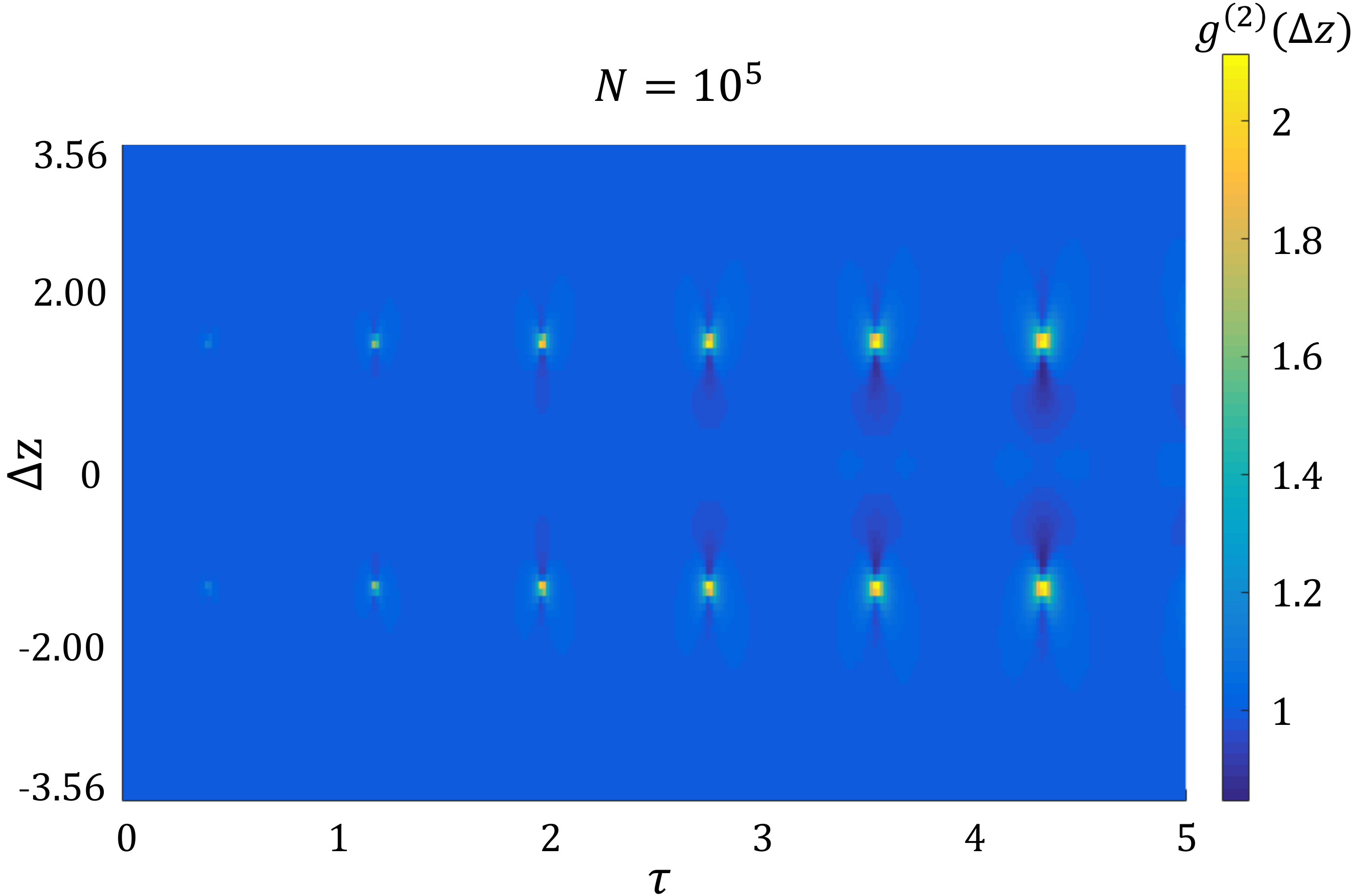}

\caption{\label{fig:g2 time evolution-10^5}Time-evolution of the normalized
second order correlation $g^{(2)}(\Delta z)$, with $N=10^{5}$. Other
parameters as in Figure \ref{fig:g2 time evolution-10^3}. }
\end{figure}

At larger $N$ values of $N=10^{5}$, the peak value of $g^{(2)}$
is significantly reduced to $\sim1.1$ at $\tau=5.0$, which appears
to give a weaker bunching within the soliton. Figures \ref{fig:g2 time evolution-10^3}
and \ref{fig:g2 time evolution-10^5} show the time evolution of the
nonlocal correlation, $g^{(2)}(\Delta z/2,-\Delta z/2)$ at different
times, for $N=10^{3}$ and $N=10^{5}$ respectively. These graphs
show that the strongest correlations occur near the peak intensities,
and are almost unchanged with particle number. What changes with $N$
is the width in time of these correlations, as they remain strong
for a much longer time with smaller particle number. The large anti-correlations
at long times show that fragmentation occurs to a highly asymmetric
output, with a much larger fragment occurring at $+z$ than at $-z$,
or vice-versa, leading to strongly negative correlations relative
to the vacuum level.

Next, we ask: are these simply classical correlations, or do they
have non-classical, quantum features? Classical correlations obey
the Cauchy-Schwarz inequality (CSI), which in terms of the second-order
correlation functions, $G^{(2)}(z_{1},z_{2})=\langle:\hat{n}(z_{1})\hat{n}(z_{2}):\rangle=\langle\hat{\psi}^{\dagger}(z_{1})\hat{\psi}^{\dagger}(z_{2})\hat{\psi}(z_{2})\hat{\psi}(z_{1})\rangle$,
is given by

\begin{eqnarray*}
G^{(2)}(z_{1},z_{2}) & \leq & \sqrt{G^{(2)}(z_{1},z_{1})G^{(2)}(z_{2},z_{2})}.
\end{eqnarray*}

One can introduce a correlation coefficient $C_{CSI}=G^{(2)}(z_{1},z_{2})/\sqrt{G^{(2)}(z_{1},z_{1})G^{(2)}(z_{2},z_{2})},$
to demonstrate that the system possesses {nonlocal fluctuations that
are stronger than any possible classical fluctuations, when $C_{CSI}>1$
\cite{glauber1963quantum,loudon1980non,reid1986violations,kheruntsyan2012violation}.}
For a system of identical bosons, the CSI is violated if the coefficient
$C_{CSI}$ is greater than unity. This violation implies that particle
entanglement {is possible to exist, \cite{wasak2014cauchy}}, leading
to the potential for tests of Bell correlation via an Ou-Mandel test
\cite{ou1988violation,rosales2014probabilistic}.

\begin{figure}[h]
\includegraphics[width=0.9\columnwidth]{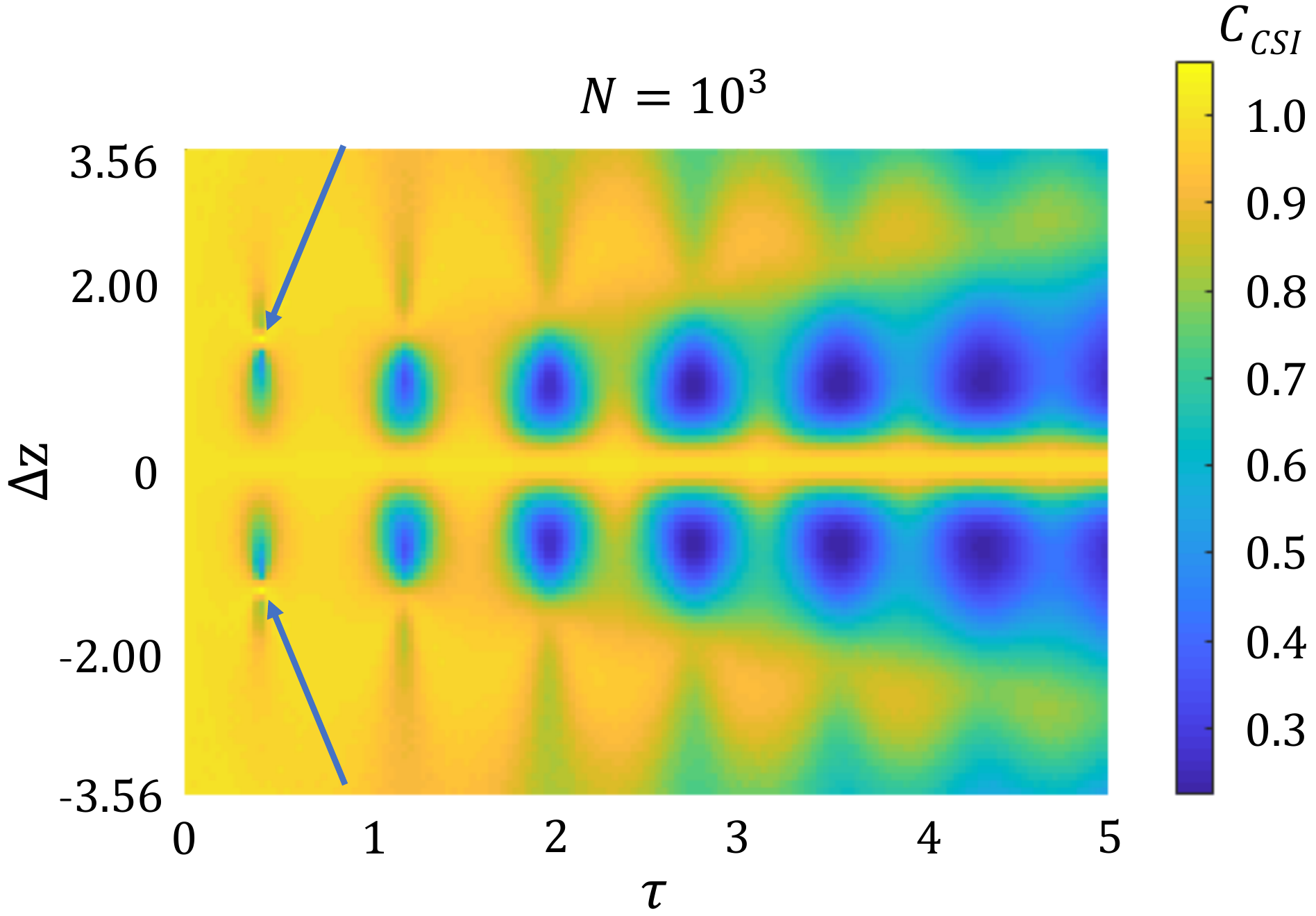}

\caption{\label{fig:coefficient contour}Time-evolution of the correlation
coefficient $C_{CSI}$. $N=10^{3}$, $CN=-8$ and $L=20$ (graph only
shows region near the centre $\Delta z=\pm3.56$). Simulations have
$M=512$ spatial modes, $8\times10^{4}$ trajectories and $3\times10^{4}$
time steps. The maximum sampling error is around $2\%$, the time-step
error is too small to be seen. Blue arrows indicate the location where
CSI is largest. }
\end{figure}
For the case of the soliton system with $1000$ particles, there is
one pair of $C_{CSI}$-peaks above unity, as shown in Figure \ref{fig:coefficient contour}.
The generation of multiple fragmented condensates at later times reduces
the strong pairing correlations found at earlier times. The peaks
which maximize the CSI correspond to the $g^{(2)}$-peaks at the same
position and the same time ($\Delta z\sim\pm1.12$, $\tau\sim\pi/8$),
as shown in Figure \ref{fig:g2 comparison-slice N1e3-1} for each
$N$. These low $N$ results have relatively large sampling errors.

\begin{figure}[h]
\includegraphics[width=0.9\columnwidth]{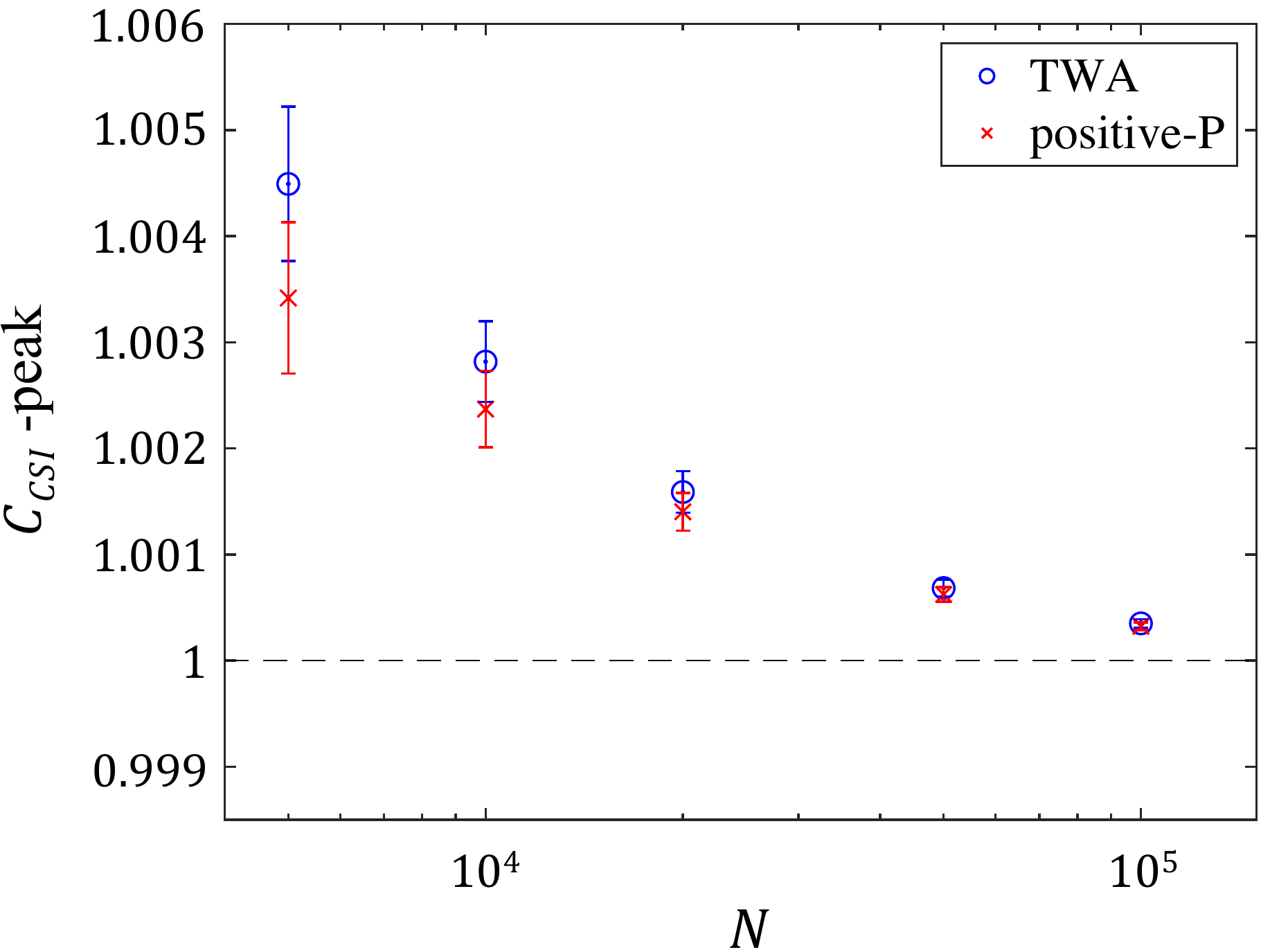}\caption{\label{fig:CSI peaks}Peak values of correlation coefficient $C_{CSI}$
for Cauchy-Schwarz inequality violations at different particle numbers
$N$. Simulations with $5\times10^{5}$ trajectories and $4500$ time
steps for a time duration $\tau=0.75$. Errors are dominated by sampling
error, which is larger for smaller $N$. Comparing calculations with
the truncated Wigner approximation and exact positive-P representations,
showing increasing agreement for higher number of particles.}
\end{figure}

Testing the CSI violation with larger numbers of particles $N$ (Figure
\ref{fig:CSI peaks}), we find that there are CSI violations at $N>1000$
. However, this reduces as $N$ increases. {As already shown in Figures
\ref{fig:g2 time evolution-10^3} and \ref{fig:g2 time evolution-10^5}},
the normalized non-local correlations $g^{(2)}(\Delta z)$ of the
BEC soliton have several peak values. Only the first pair of $g^{(2)}$-peaks
gives a CSI violation. Although the $g^{(2)}$-peak values remain
strong at later time $\tau>\pi/8$ {for systems with larger number
of particles}, no further CSI violations are observed to the accuracy
of the present simulations. The soliton fluctuations apparently become
more classical as the number of particles in the soliton increases,
even though the peak intensity correlations remain very strong.

To summarize, our quantum dynamical calculations predict very strong
nonlocal anti-correlations in 1D BEC soliton breathers as they fragment.
The oscillatory decay of the nonlocal correlation depends on the particle
number $N$, with the position and the amplitude of the correlation
peak being relatively stable at large $N$, but with a reduced peak
width. There is also a small Cauchy-Schwarz inequality violation,
showing that these nonlocal correlations have nonclassical entanglement.
This effect is most pronounced at the first peak, and in a system
of mesoscopic scale ($N\sim10^{3}$). We interpret this as a radiation
of entangled, correlated pairs of particles at the soliton peaks,
which occurs during soliton fragmentation into asymmetric fragments.
At subsequent peaks the strong correlation remains. However, the localized
Cauchy-Schwarz inequality violation becomes less pronounced as the
quantum many-body system becomes more fragmented during the decay.
\begin{acknowledgments}
PDD and MDR thank the Australian Research Council and the hospitality
of the Institute for Atomic and Molecular Physics (ITAMP) at Harvard
University, supported by the NSF. PDD acknowledges useful discussions
with M. Olshanii. This research has been supported by the Australian
Research Council Discovery Project Grants schemes under Grant DP180102470.
\end{acknowledgments}

 \bibliographystyle{apsrev4-1}

%

\end{document}